\documentclass[aps, showpacs, prb,superscriptaddress,twocolumn,a4paper]{revtex4-1}
\usepackage{graphicx}
\usepackage{amsmath, amsfonts, amssymb, bm}
\usepackage{color}
\usepackage{multirow}
\usepackage[utf8]{inputenc}
\usepackage[english]{babel}
\usepackage{amsmath,array,amssymb,pst-all}
\usepackage{dsfont}
\usepackage{bbold}

\newcommand{\sugb}[1]{\textcolor{blue}{#1}}

\usepackage[para,online,flushleft]{threeparttable}

\begin{document}
\title{Theoretical Characterization of Photoactive Molecular Systems Based on BODIPY-Derivatives for the Design of Organic Solar Cells}

\author{Duvalier Madrid-Úsuga}
\altaffiliation{{\tt{duvalier.madrid@correounivalle.edu.co}}}
\address{Centre for Bioinformatics and Photonics---CIBioFi, Universidad del Valle, Edificio 320, No. 1069, 760032 Cali, Colombia}
\address{Quantum Technologies, Information and Complexity Group---QuanTIC, Department of Physic, Universidad del Valle, 760032 Cali, Colombia}

\author{Ana G. Mora-Leon}
\address{Engineering and Environmental Management Research Group---GIGA, School of Engineering, Universidad de Antioquia, 050010 Medellin, Colombia}

\author{Andrea Cabrera-Espinosa}
\address{Centre for Bioinformatics and Photonics---CIBioFi, Universidad del Valle, Edificio 320, No. 1069, 760032 Cali, Colombia}
\address{Heterocyclic Compounds Research Group---GICH, Department of Chemistry, Universidad del Valle,  760032 Cali, Colombia}

\author{Braulio Insuasty}
\address{Centre for Bioinformatics and Photonics---CIBioFi, Universidad del Valle, Edificio 320, No. 1069, 760032 Cali, Colombia}
\address{Heterocyclic Compounds Research Group---GICH, Department of Chemistry, Universidad del Valle,  760032 Cali, Colombia}

\author{Alejandro Ortiz}
\altaffiliation{{\tt{alejandro.ortiz@correounivalle.edu.co}}}
\address{Centre for Bioinformatics and Photonics---CIBioFi, Universidad del Valle, Edificio 320, No. 1069, 760032 Cali, Colombia}
\address{Heterocyclic Compounds Research Group---GICH, Department of Chemistry, Universidad del Valle,  760032 Cali, Colombia}

  
\begin{abstract}
To search for high-efficiency narrow-band donor materials to improve the short-circuit current density ($J_{sc}$) of organic solar cells, a series of small molecules based in Bodipy-Triphenylamine were characterized using density functional theory (DFT) and time-dependent (TD-DFT) calculations. According to the energy of the exciton driving force they have the appropriate energy levels to match \textbf{\textit{PC$_{61}$BM}}. The properties affecting the open circuit voltage ($V_{oc}$), $J_{sc}$ and the fill factor ($FF$) were investigated by calculating the geometric structures, the boundary molecular orbital energy levels, absorption spectra, light collection efficiencies, chare transfer rates, and exciton binding energies. The results show that the \textbf{\textit{BTPA~III}} system has a lower LUMO level, high absorption efficiency, and exction dissociation than other molecular systems, facilitating the improvement of $V_{oc}$, $J_{sc}$ and $FF$. Finally, \textbf{\textit{BTPA~III}} would be the most promising of this series of donors and further increase the efficiency of the device.\\
\\

\end{abstract}

\maketitle
\section{Introduction}

Organic Solar Cells (OSC) of bulk heterojunption (BHJ) consist of a mixture of donor molecular systems and/or polymer and electron transporting material (ETM), generally derived from fullerene. These cells have attracted great attention from the physical, chemical and research fields due to its various advantages over traditional silicon solar cells, such as  low cost, light weight, easy manufacturing, transparency and flexibility~\sugb{\cite{you2013, huang2014, aruna2015, wang2016, xia2017}}.  Although good results have been obtained using different systems such as shown by Yang et al., who reported a $PCE=6.0\% $ using gasymmetrical squaraines (ASQs) donor molecules electron~\sugb{\cite{yang2015}}, or those reported by Ni et al .;  Collins et al., with a $PCE=10.0 \%$~\sugb{\cite{ni2015, collins2017}}, or for the case where obtained results using fullerene-free electron-acceptor materials with a $PCE$ around $14.0 \%$~\sugb{\cite{zhang2018, zhang2018, yan2018}}. Still challenging to improve energy conversion efficiency ($PCE$) in view of the intrinsic drawbacks of the molecular systems that make up the active layer, including light absorption in the visible region of the solar radiation, strong tendency to molecular aggregation, difficult to adjust energy levels through molecular alteration, among other things.
\medskip
To improve the eficiency of OSCs, materials should be designed carefully to tune the frontier molecular orbitals (FMOs). For on the other hand, the HOMO (highest occupied molecular orbital) energy level of the donor molecular systems must be reduced to obtain a voltage of large open circuit ($V_{oc}$).  On the other hand, the LUMO energy level (orbital lowest unoccupied molecular weight) of the donor must be at least $0.3$~eV plus higher than the acceptor's LUMO level (for example, \textbf{\textit{$PC_{61}BM $}}) to ensure quantitative separation of charge at the interface D/A~\sugb{\cite{zhang2012, kim2013}}.
\medskip
On the other hand, theoretical knowledge of the HOMO and LUMO energy levels, gap of energy, and spectrum of absorption of molecular complexes is fundamental in the study of OSCs, so today, theoretical quantum calculations have been effective tools in the field of physico-chemistry because they can be used to rationalize these properties of known chemical compounds and also model those of unknown compounds to guide experimental synthesis observed~\sugb{\cite{Belghiti2012}}. The computational tools of the quantum chemistry has provided many solutions to the questions posed by experimentalists, namely reaction mechanisms, the calculus of structural parameters (particularly in the case of amorphous products), the modeling and prediction of current properties between hole transporting meterial (HTM) and electron transporting material (ETM), in addition to determining the optical and photovoltaic electronic properties. These methods are based on approximations; in particular, the density functional theory (DFT) has been very developed and takes into account the important effects of electronic correlation. To Unlike Hartree Fock (HF) methods, they can provide reliable reliability very good for certain molecular systems, especially conjugates.
In this work, the electronic structure and the optical absorption properties, thus as the photovoltaic properties of four derivatives of BODIPY and triarylamine they were calculated using DFT and TDDFT. Based on the calculated results, we analyze the role of different electron donor groups in adjusting the geometries, electronic structures and optical properties. In addition, our The objective was to see the effects of the molecular structures of the hole transporting materials on the open circuit photovoltage ($V_{oc}$), the density of short-circuit current ($J_{sc}$) and the PCE of the organic solar cell of BHJ. For the which, the Scharbers model is implemented, which although it cannot be use by itself to make precise predictions of the $ PCE $, since the values obtained from $J_{sc}$ do not constitute a strict upper limit, since the model uses very simplistic assumptions for the $EQE$ value, which is a complicated frequency-dependent function that would have to take into account transport effects and device morphology, and therefore $J_{sc}$ should only be considered as an indication of molecular systems potential that could be harvested if the device possesses a good the external quantum efficiency ($EQE$)~\sugb{\cite{berube2013}}. However, the Scharbers model together with theoretical calculations based on Kohn-Sham energy levels from density functional theory can provide a guide to finding promising molecular and/or polymer systems for photovoltaic organic solar cells, indicating that the model provides an important guide for future research and that it has been widely used in the past and today to measure the potential in the application of solar cells of different molecular systems~\sugb{\cite{scharber2006, dennler2009, azazi2011, scharber2013, berube2013, wang2014, scharber2016, taouali2019, siddiqui2019, padula2019, raftani2020, afzal2020, hachi2020, mahdavifar2020}}.
\medskip
This paper is organized as follows: molecular systems and details computational data are shown in Sec.~\sugb{\ref{computational}};  the absorption spectra and electron contributions are shown in the Sec.~\sugb{\ref{Absorption}}, the energies of HOMO, LUMO, the energy and density of state (DOS) of molecular systems are shown in the  Sec~\sugb{\ref{Electronic}}. the main results of the properties photovoltaics, as well as the power conversion energy (PCE) are shown in the Sec.~\sugb{\ref{Photovoltaic}}.  Finally, conclusions are given in Sec.~\sugb{\ref{Conclusions}}.

\section{Results and Discussion} \label{sec:result}

\subsection{Molecular Systems and Computacional Detail} \label{computational}

\begin{figure}[htp]
\includegraphics[scale=0.31]{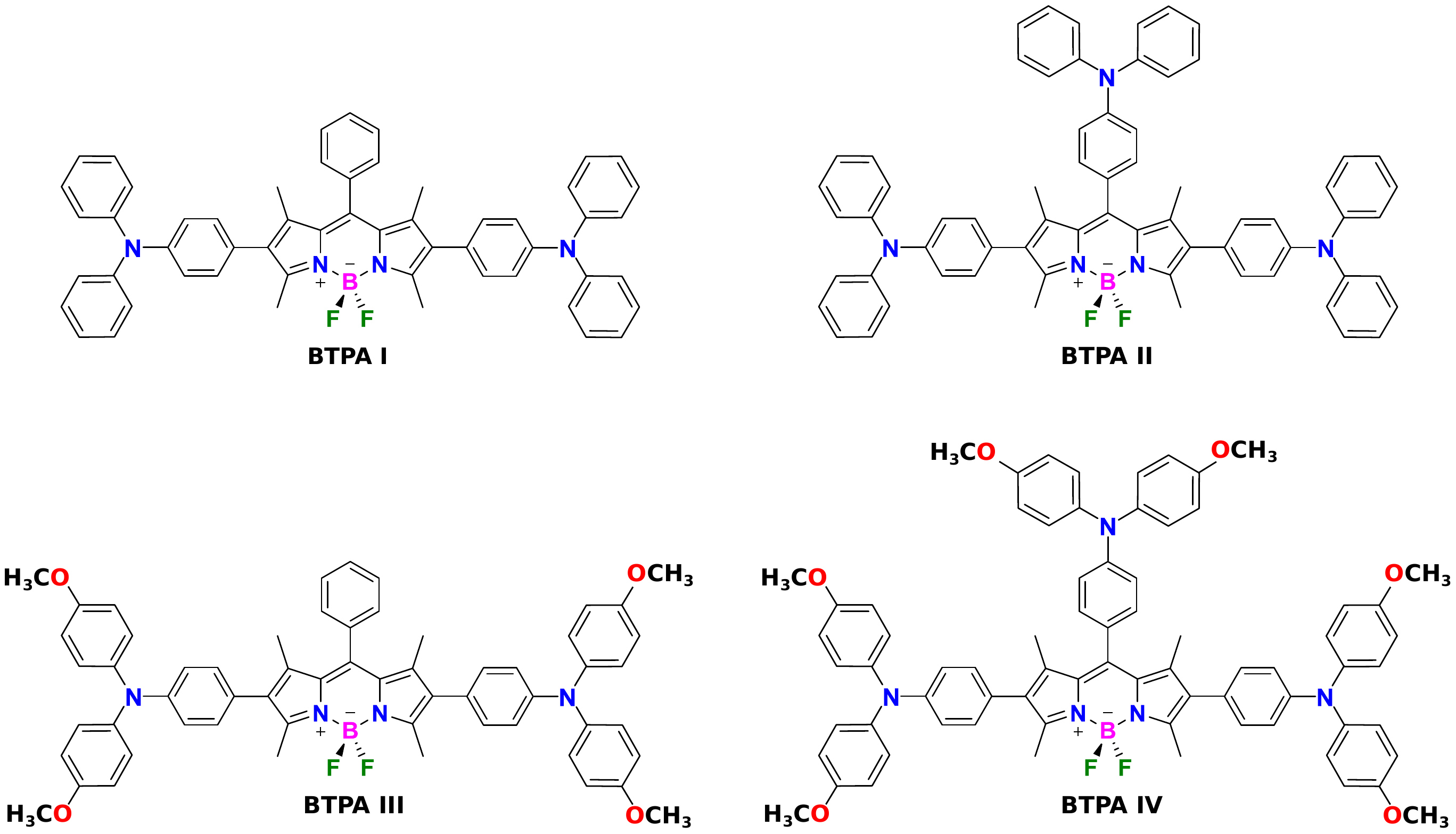}
\caption{Optimized structures of the molecular compounds \textbf{\textit{BTPA~I}}, \textbf{\textit{BTPA~II}}, \textbf{\textit{BTPA~III}}, \textbf{\textit{BTPA~IV}}.}
\label{Fig1}
\end{figure}
With the intention of contributing to the development of electron-donor materials for the active layer of solar cells, this work considers a computational study of electronic, optical and photovoltaic properties of the molecular systems \textbf{\textit{BTPA~I}}, \textbf{\textit{BTPA~II}}, \textbf{\textit{BTPA~III}}, \textbf{\textit{BTPA~IV}} presented in Fig.~\sugb{\ref{Fig1}}, whose synthesis was previously reported by A. Ortiz~\sugb{\cite{ortiz2019}}. All the calculations were carried out using Gaussian 09 Package~\sugb{\cite{frisch2009}}. 

The geometries obtained for such most stable conformations (see supplementary section) were used as input data for the full optimization of calculations of the ground state by means of the density functional theory (DFT) with B3LYP (Becker Three-parameter Lee-Tang-Parr) exchange-correlation~\sugb{\cite{becke1993}} and the base set 6-31G(d,p) to the optimization geometry of the ground state of the systems and predict the energies of frontier molecular orbital (FMO)~\sugb{\cite{madrid2018,calderon2020}}. The vertical excitation energies, the absorption spectrum, the oscillator strengths, and the transition dipole moment are based on the TD-DFT, using the based set CAM-B3LYP which has demonstrated its ability to predict the excitation energy and the absorption spectral of different molecules~\sugb{\cite{ganji2015,ganji2016}}. The influence of the molecular environment was modeled as a dielectric medium employing a conductor-like polarization continuum model (C-PCM)~\sugb{\cite{takano2005, chiu2016}}. The C-PCM is regularly used to recognize the solvation which indicates the solvent effects in the molecular complex. In the present study, we consider as dissolvent the Tetrahydrofuran (THF) represented by the dielectric constants $\varepsilon_{_{THF}}=7.4257$ to follow the polarity effects on the molecular photo-physics and to be able to compare some of the results with those obtained experimentally.

\subsection{Absorption Properties}\label{Absorption}

\begin{figure}[htp]
\centering
\includegraphics[scale=0.58]{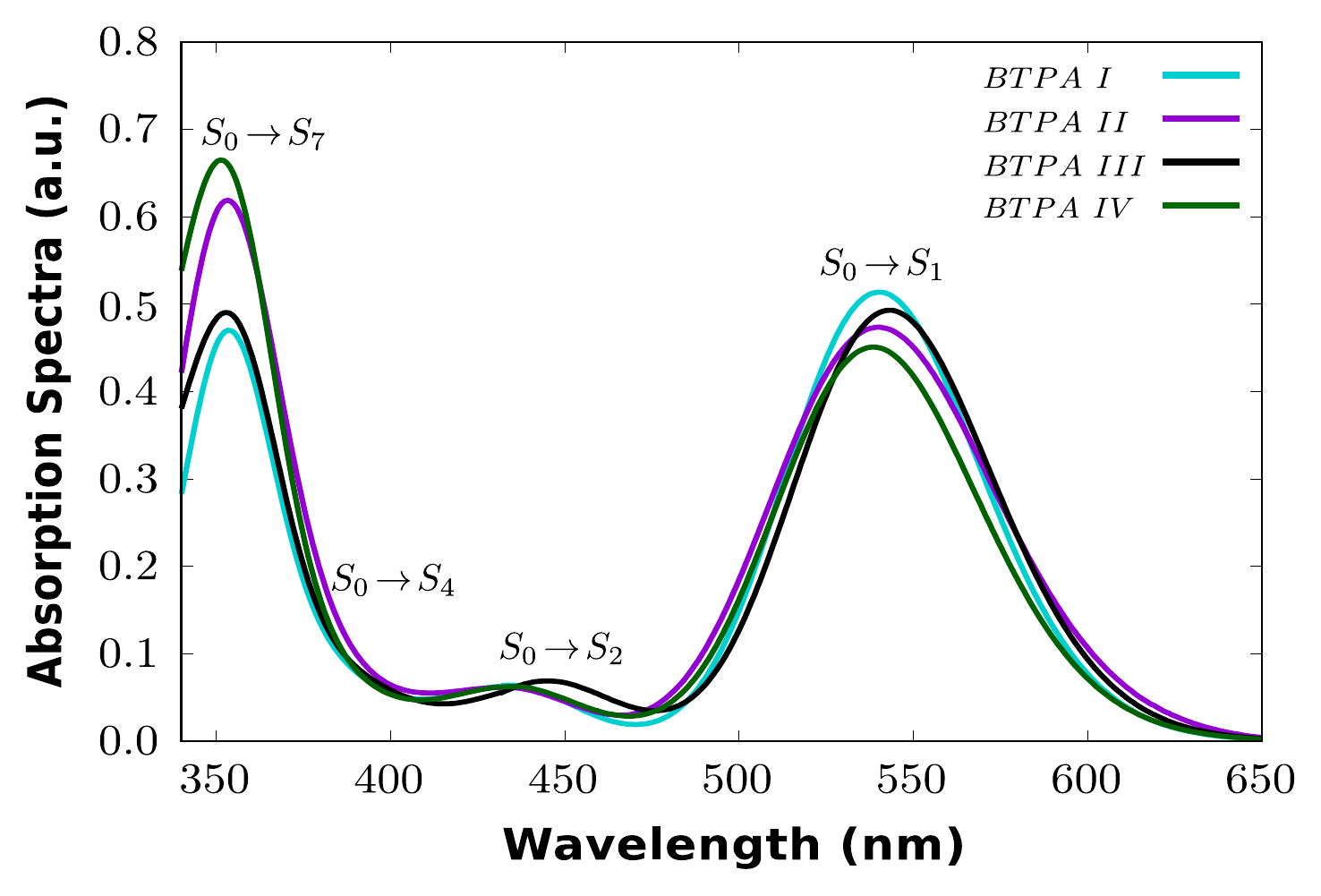}
\caption{Simulated absorption spectra of derivatives \textbf{\textit{BTPA~I}}, \textbf{\textit{BTPA~II}}, \textbf{\textit{BTPA~III}}, \textbf{\textit{BTPA~IV}} in the presence of Tetrahidrofurano solvent, using the TD-DFT/CAM-B3LYP/6-31G(d,P) level of theory.}
\label{Fig2}
\end{figure}

The various optical properties such as the excitation energies and UV/Vis absorption spectra for the dyes, \textbf{\textit{BTPA~I--IV}}, were calculated in THF using TD-DFT with CAM-B3LYP functional. The optimized structures of the \textbf{\textit{BTPA~I--IV}} compounds are shown in Fig.~\sugb{\ref{Fig1}} and the spectra are shown in Fig.~\sugb{\ref{Fig2}}, which have good agreement with those obtained experimentally. We computed vertical excited singlet states, transition energies, oscillator strengths and the maximum absortion length $\lambda_{max}$ of the dye derivatives, corresponding to first 20 singlet excited states in THF, and the results are tabulated in Table~\sugb{\ref{Tab1}}. The spectra show that for the derivatives of \textbf{\textit{BTPA~I}} and \textbf{\textit{BTPA~II}} show a strong absorption around $539.090$~nm and $536.910$~nm respectively, corresponding to the transition $S_0\rightarrow S_1$, associated with the electronic transition HOMO $\rightarrow$ LUMO with a contribution of $53\%$ in both systems. 
\medskip

\begin{figure*}[htp]
\includegraphics[scale=0.6]{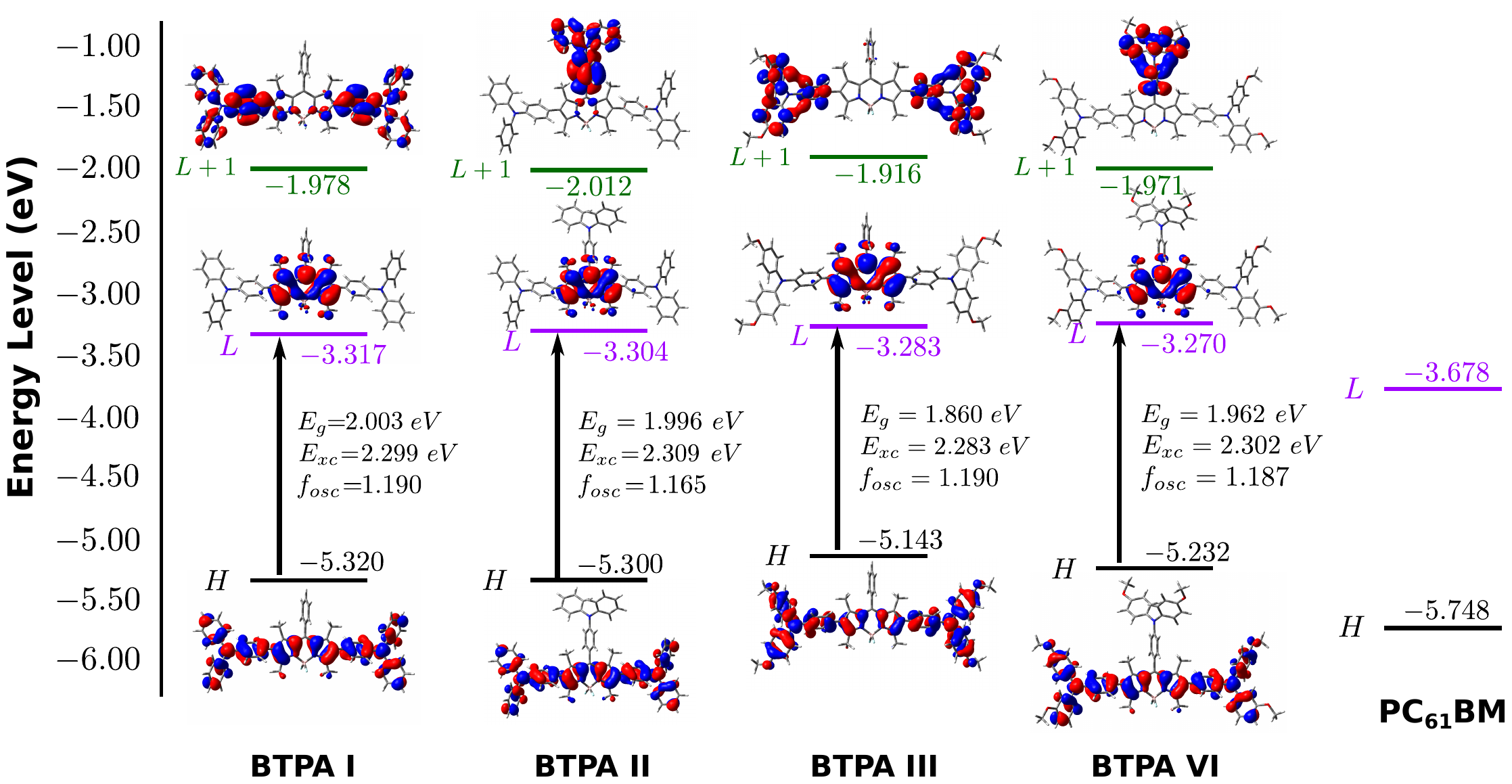}
\caption{Electron density maps of the frontier molecular orbital HOMO and LUMO of \textbf{\textit{BTPA~I}}, \textbf{\textit{BTPA~II}}, \textbf{\textit{BTPA~III}}, \textbf{\textit{BTPA~IV}} in the presence of THF.}
\label{Fig3}
\end{figure*}

The \textbf{\textit{BTPA~I}} system has an oscillator force of $f_{osc}=1.190$ and the \textbf{\textit{BTPA~II}} system has$f_{osc}=1.165$. On the other hand, for the compound \textbf{\textit{BTPA~III}}, the optically allowed electronic transition is related to populate the HOMO $\rightarrow $ LUMO excitation with high oscillator strength $f_{osc}=1.190$, which is related to a high energy band that registers an absorption peak at $543.110$~nm in the absorption spectrum Fig.~\sugb{\ref{Fig2}}, which can be ascribed to the intramolecular charge transfer in the BODIPY part. While the \textbf{\textit{BTPA~IV}} system presents a low energy absorption band at $538.470$~nm, with a $f_{osc}=~1.187$ associated with the electronic transition HOMO $\rightarrow$ LUMO ($48\%$). The electronic transitions of the molecular systems are presented in Fig.~\sugb{\ref{Fig3}}. In addition, the systems present a small high energy band around $ 440.250$~nm that exhibits a very small oscillator force compared to the $f_{osc}$ of the maximum absorption band.

\begin{table}[htp] 
\begin{center}
\centering
\caption{Maximum absorption wavelength $\lambda_{max}$, the excitation energy \textit{E$_{exc}$}, the oscillator strength \textit{f$_{osc}$}, and the contribution of the most probable transition of the studied compounds.}
\scalebox{0.8}{
\begin{tabular}{ccccccc} 
\hline
\hline
&&&&\\
\centering \textbf{\textit{System}} & \textbf{\textbf{\textit{E}}$_{xc}$ (eV)} & \textbf{$\lambda_{max}$ (nm)} & \textbf{\textit{f$_{osc}$}} & \textbf{\textit{LHE}} & \textbf{\textit{Composition}} \\
&&&&\\
\hline
\hline
&&&&\\
\textbf{\textit{BTPA~I}}   & 2.299 & 539.090 & 1.191 & 0.936 & $H \rightarrow L$ (53\%) \\
\textbf{\textit{BTPA~II}}  & 2.309 & 536.910 & 1.165 & 0.932 & $H \rightarrow L$ (53\%) \\
\textbf{\textit{BTPA~III}} & 2.283 & 543.110 & 1.190 & 0.935 & $H \rightarrow L$ (51\%) \\ 
\textbf{\textit{BTPA~IV}}  & 2.302 & 538.470 & 1.187 & 0.936 & $H \rightarrow L$ (48\%) \\  
&&&&\\
\hline
\hline
\end{tabular}}
\label{Tab1}
\end{center}
\end{table}

On the other hand, the change in the absorption maxima towards the region of greater wavelength is beneficial for a better collection of solar radiation and, therefore, improve the efficiency of the system. Therefore, we observe that the inclusion of triarylamine (TPA) groups in the structure of \textbf{\textit{BTPA~II}} compared to \textbf{\textit{BTPA~I}}, decreases the absorption range generated by the molecular system, however, the inclusion of the methoxyl group improves the absorption rate of the molecular system \textbf{\textit{BTPA~III}}, favoring the efficiency in the collection of radiation. Which improves its applicability as an electron donor material in the active layer of a BHJ organic solar cell.
On the other hand, the oscillator force associated with the wavelength of maximum absorption, gives us information on the light capture efficiency (\textit{LHE}) of each molecular compound when they are exposed to solar radiation, which can be calculated from the equation~\sugb{\cite{zhang2013}},
\begin{equation}
LHE=1-10^{-f_{osc}}
\label{Ecu1}
\end{equation}
The \textit{LHE} is an important factor in OSCs, since this value gives an intuitive impression of the efficiency of the electron-donor compound in the photon absorption process or of the total absorptivity of the molecule. According to Equ.~\sugb{\ref{Ecu1}}, higher values of $f_{osc}$ are reflected in higher $LHE$. The highest value of $LHE$ is, in addition to other factors, a representative of the depleted photocurrent response of the molecule. The Table~\sugb{\ref{Tab1}} shows the LHE values of molecular compounds, which are in the range $0.932$--$0.936$. This means that all compounds have a similar sensitivity to sunlight. The $LHE$ of the molecular derivatives under discussion follows the order: \textbf{\textit{BTPA~II}}$<$\textbf{\textit{BTPA~IV}}$<$\textbf{\textit{BTPA~I}}$<$\textbf{\textit{BTPA~III}}. Considering both $\lambda_{max}$ and $LHE$, \textbf{\textit{BTPA~III}} might have the higher absorption efficiency.
\medskip

To understand the origin of the absorption spectra of different dyes, we record the isodensity plots of the boundary molecular orbitals involved in the dominant transitions (see Fig.~\sugb{\ref{Fig3}}). The electron density distribution in the isodensity plots supports the view that these transitions involve the transfer of charge from the triarylamine groups to the BODIPY (acceptor). This intramolecular charge transfer leads to absorptions in the higher wavelength regions and therefore leads to better collection of solar radiation. The optimized geometry of the molecular systems and their HOMO and LUMO energy levels are shown in Fig.~\sugb{\ref{Fig3}}. The geometry indicates that the HOMO orbital has scattered the electron cloud over the entire TPA-BODIPY-TPA unit. While in the LUMO orbital, the electron cloud has spread to the BODIPY electron acceptor unit from the TPA electron donor unit. The result of the shift of the electron cloud from the TPA units to the electron acceptor unit in the BODIPY indicates the existence of a strong intramolecular charge transfer.

\subsection{Electronic Properties}\label{Electronic}

OSCs based on a BHJ structure generally represent a mixture of an electron-donor $\pi$--conjugated material with an electron-acceptor material derived from fullerene. In this work, [6,6]-phenyl-C$_{61}$-butyric acid methyl ester (\textbf{\textit{PC$_{61}$BM}}) will be considered as the electron acceptor material in the active layer, which is a widely used electron acceptor in solar cell related devices, as it has excellent electron transport combined with LUMO energy that is high enough to withstand large photovoltage, but also low enough to provide ohmic contacts for the extraction and injection of electrons from common cathode electrodes~\sugb{\cite{jagadamma2015}}. Knowing theoretically the HOMO and LUMO energy levels of these components is crucial in the study of the properties associated with solar cells, since it allows to know if the effective charge transfer will occur between the electron-donor material and the electron-acceptor within the active layer of organic solar cells. In OSCs, the energy of the HOMO-LUMO boundary orbitals associated with the materials of the active layer has a direct relationship with the photovoltaic properties, since these levels are related to the open circuit ($V_{oc}$) and the energy driving force ($\Delta E$) for exciton dissociation~\sugb{\cite{brabec2011}}. To estimate the energy of the HOMO-LUMO boundary orbitals, the DFT level of theory and the fitted equations of Bérubé et al., expressed as~\sugb{\cite{berube2013}} were implemented:

\begin{eqnarray}
E_{H} &=& 0.68\times E_{H_{_{DFT}}}-1.92 \; {\textrm{eV}} \\ \nonumber
\label{Equ2}
E_{L} &=& 0.68\times E_{L_{_{DFT}}}-1.59 \;  {\textrm{eV}}
\end{eqnarray}
\begin{table*}[htp] 
\begin{center}
\centering
\caption{Calculated values of the frontier orbitals HOMO (H) and LUMO (L) energies which constitute the photo-activated material with the respective estimation (\textit{$E_{_{H}}$} and \textit{$E_{_{L}}$}), the energy gap $E_{g}$, optical energy $E_{opt}$, binding energy $E_{b}$ and the energy of the exciton driving force $\Delta E$ (all in eV) for exciton dissociation in THF.}
\scalebox{0.85}{
\begin{tabular}{ccccccccc} 
\hline
\hline
\\  
\multicolumn{1}{m{1.5cm}}{\centering \textbf{\textit{System}}} & \multicolumn{1}{m{1.8cm}}{\centering\textbf{\textit{E$_{_{H_{DFT}}}$}}(eV)} & \multicolumn{1}{m{1.5cm}}{\centering\textbf{\textit{E$_{_{L_{DFT}}}$}}(eV)} & \multicolumn{1}{m{1.5cm}}{\centering\textbf{\textit{E$_{_{H}}$}}(eV)} & \multicolumn{1}{m{1.5cm}}{\centering\textbf{\textit{E$_{_{L}}$}}(eV)} & \multicolumn{1}{m{1.5cm}}{\centering\textbf{\textit{E}$_{g}$}(eV)} & \multicolumn{1}{m{1.5cm}}{\centering\textbf{\textit{E}$_{opt}$}(eV)} & \multicolumn{1}{m{1.5cm}}{\centering\textbf{\textit{E}$_{b}$}(eV)} & \multicolumn{1}{m{1.5cm}}{\centering\textbf{$\Delta$\textit{E}}(eV)} \\
&&&&&&\\
\hline
\hline
&&&&&&\\
\textbf{\textit{BTPA~I}}       & -5.000 & -2.540 & -5.320 & -3.317 & 2.003 & 1.813 & 0.190 & 0.360 \\
\textbf{\textit{BTPA~II}}      & -4.990 & -2.520 & -5.313 & -3.304 & 2.010 & 1.830 & 0.180 & 0.370 \\
\textbf{\textit{BTPA~III}}     & -4.740 & -2.490 & -5.143 & -3.338 & 1.806 & 1.716 & 0.090 & 0.340 \\ 
\textbf{\textit{BTPA~IV}}      & -4.780 & -2.470 & -5.170 & -3.270 & 1.901 & 1.751 & 0.150 & 0.408 \\  
\textbf{\textit{PC$_{61}$BM}}  & -5.630 & -3.070 & -5.748 & -3.678 & 2.071 &  ---  &  ---  &  N.A  \\
&&&&&&\\
\hline
\hline
\end{tabular}}
\label{Tab2}
\end{center}
\end{table*}
The results of the HOMO/LUMO values calculated using DFT and the Equ.~\sugb{2} of the molecular systems~\textbf{\textit{BTPA~I--IV}}, as well as the electron material-acceptor in the active layer \textbf{\textit{PC$_{61}$BM}} are shown in Table~\sugb{\ref{Tab2}}. For molecular systems with application in OSCs, the energy of the boundary orbitals of the donor and acceptor molecules are an important issue to consider, because they are considerably relative to the open circuit voltage ($V_{oc}$) and the force driving energy ($\Delta E$) for exciton dissociation. The value of $V_{oc}$ can be determined by the difference in orbital energy between the $E_{_H}$ of the electron donor molecule (\textbf{\textit{BTPA~I--IV}}) and the $E_{L}$ of the electron acceptor molecule \textbf{\textit{PC$_{61}$BM}}~\sugb{\cite{scharber2006, jo2012, wang2014, wang2014r}}, and the energy difference, $\Delta E$, can be estimated by the difference of $E_{L}$ between donor and electron acceptor molecules \textbf{\textit{PC$_{61}$BM}}, which should be large than $0.3$~eV to ensure exciton separation and charge dissociation in the electron donor/electron acceptor interface~\sugb{\cite{walker2013, duan2013, wang2014}}. A molecule with a large $\Delta E$ shows that an effect of exciton separation and charge dissociation can be generated at the interface of the electron donor and the electron acceptor.
\medskip

To analyze these properties, Table~\sugb{\ref{Tab2}} lists $E_H$, $E_L$, and $E_g$ of the molecular systems and the Fig.~\sugb{\ref{Fig3}} shows the energy diagram of the boundary molecular orbitals (FMO) of all the hole transporting material with respect to the electron transporting material \textbf{\textit{PC$_{61}$BM}}. The calculated $E_{H}$ exhibits a tendency to increase for the derivatives \textbf{\textit{BTPA~II}}, \textbf{\textit{III}} and \textbf{\textit{VI}} with different structures molecular, with respect to the \textbf{\textit{BTPA~I}} ($-5.00$~eV) system. It can be seen that the calculated energy levels of LUMO and HOMO of \textbf{\textit{BTPA~II}}, \textbf{\textit{III}} and \textbf{\textit{VI}}, respectively grow about $0.013$--$0.047$, $0.010$--$0.177$, while and the band gaps drop about $0.041$--$0.088$~eV compared to \textbf{\textit{BTPA~I}}, which implies that inclusion of triarylamine and the methoxy group clearly increases the LUMO and HOMO energy levels, and thus lessens the band gaps, potentially enhancing the $V_{oc}$ and $J_{sc}$. Furthermore, a D-$\pi$-A molecule with a longer conjugated length $\pi$-exhibits a lower $E_g$, which guarantees a higher conjugation in the \textbf{\textit{BTPA~III}} system followed by the molecular system \textbf{\textit{BTPA~VI}}. We also observe that all systems show acceptable values of $\Delta E$, but mainly the systems \textbf{\textit{BTPA~I}} and \textbf{\textit{III}}, which have the lowest values of $\Delta E $, which indicates that these systems ensure effective exciton dissociation and reduce energy loss more than the \textbf{\textit{BTPA~III}} and \textbf{\textit{BTPA~IV}} system. This is of great importance to our systems, since molecular systems with low $\Delta E$ and high $V_{oc}$ should improve the performance of a solar cell.

The binding energy is another promising factor for the evaluation of optoelectronic properties. A lower bond energy ($E_b$ helps to destroy the coulomb attraction between the hole and the electron and facilitates the dissociation of the exciton. The bond energy is estimated by the difference in the HOMO-LUMO energy gap ($E_g$) and the first singlet exciton energy $E_{opt}$. These values represent the exciton transfer from the $S_0$ (ground) to $S_1$ (excited) state to give a hole-electron pair. In general , $E_b$ can be expressed using the following relation~\sugb{\cite{dkhissi2011, kose2012}}

\begin{equation}
E_b=E_{H-L}-E_{opt}
\label{Ecu3}
\end{equation}

The calculated results are shown in Table~\sugb{\ref{Tab2}}, it can be seen that all the molecules show small $E_b$ values, which are less than $0.2$~eV. The $E_b$ values of the designed molecules can be ordered as follows \textbf{\textit{BTPA~III}}$<$\textbf{\textit{BTPA~IV}}$<$\textbf{\textit{BTPA~II}}$<$\textbf{\textit{BTPA~I}}. Consequently, the combination of the analysis of charge transfer and exciton binding energy could more effectively promote exciton separation and potentially improve $J_{sc}$. Since \textbf{\textit{BTPA~I--IV}} molecules have a low link energy, comparable to molecular systems used in solar cells~\sugb{\cite{clarke2010}}, all the molecules studied can show greater dissociation which results in enhancement of overall current charge density. Furthermore, all the investigated molecules (\textbf{\textit{BTPA~I--IV}}) show a low binding energy which improves the exciton dissociation efficiency in the excited state by illustrating charge separation more efficiently.
\medskip

\begin{figure}[htp]
\includegraphics[scale=0.7]{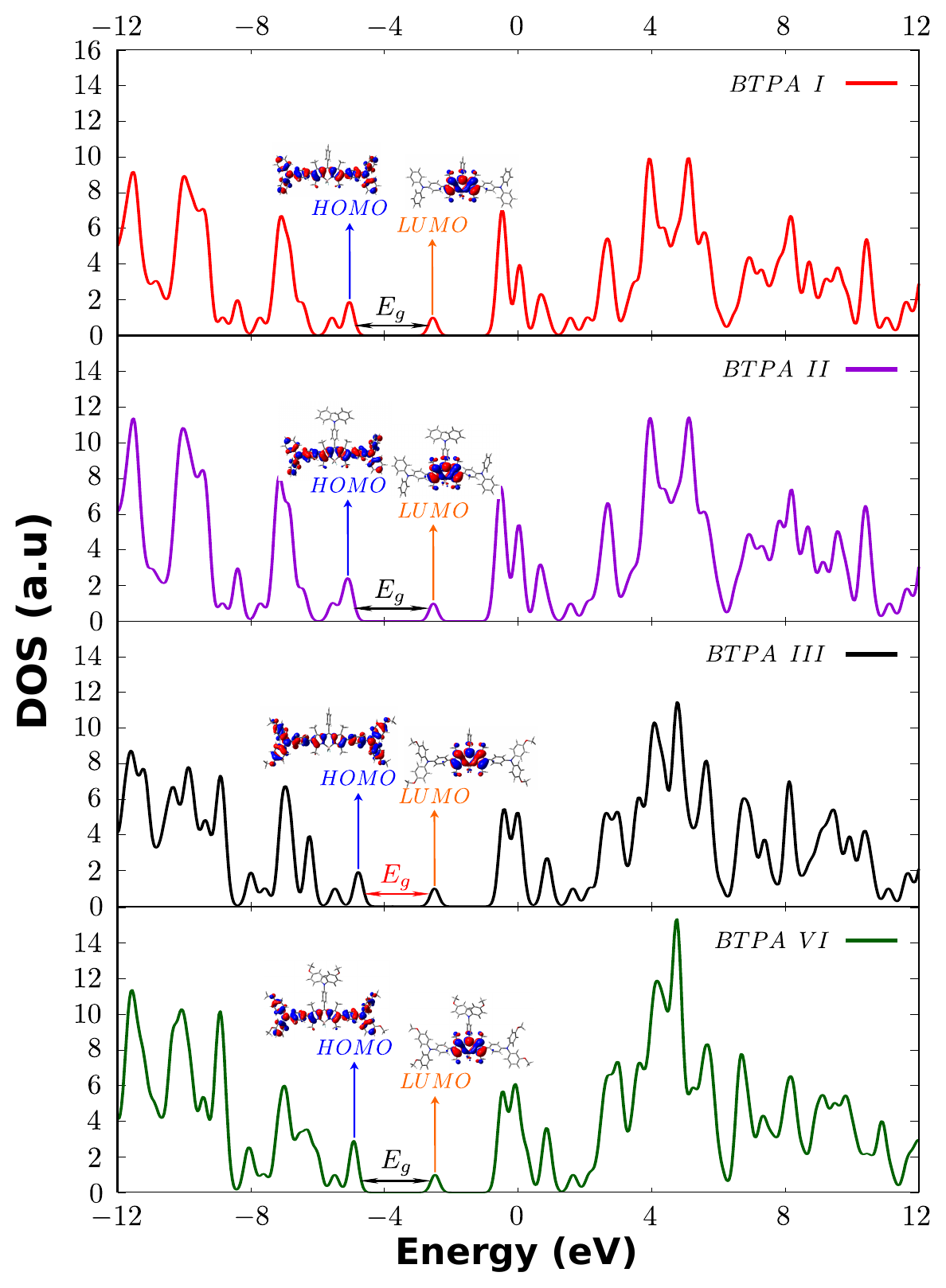}
\caption{HOMO, LUMO energies and density of states (DOS) for all the molecular systems of study.}
\label{Fig3a}
\end{figure}
On the other hand, many types of research have shown that the study of density of states (DOS) gives adequate explanations about the properties of charge transport, specifically in the case of semiconductor materials. The DOS of the system defines the number of states contained in an energy interval for each energy level that are exposed to be occupied by electrons~\sugb{\cite{renuga2014}}. The density distributions are not discrete like a spectral density, but they are infinite. The high DOS at specific power levels shows many states that are available for occupancy. The DOS show a high delocalization of the energy levels for the molecules studied as a result of the simultaneous response of the acceptor and donor group, which can be observed in Fig.~\sugb{\ref{Fig3a}}. The energy of the band gap increases in parallel with the density of states (DOS) of the compounds studied from the acceptor and donor parts of the D-$\pi$-A structure. This is explained by the effect of the energetic disorder on the transport of electrons and holes, which remains stable if the transport is determined by the main body of DOS.

The energy band gap that separates the highest occupied and the lowest unoccupied molecular orbitals characterizes the molecular electrical transport properties through the measurement of electronic conductivity~\sugb{\citep{dhas2010}}. The $E_g$ de results are $2.003 $, $2.010$, $1.806$ and $1.901$~eV for \textbf{\textit{BTPA~I}}, \textbf{\textit{II}}, \textbf{\textit{III}} and \textbf{\textit{IV}} respectively. These low values define the eventual charge transfer interactions that take place within~\sugb{\cite{sangeetha2016}} molecules. Thus, the band gap energy increases in parallel with the density of states (DOS) of the compounds studied from the acceptor and donor parts of the D-$\pi$-A structure. This is explained by the effect of the energetic disorder on the transport of electrons and holes, which remains stable if the transport is determined by the main body of the DOS.

\subsection{Photovoltaic properties}\label{Photovoltaic}

\begin{figure}[htp]
\centering
\includegraphics[scale=0.6]{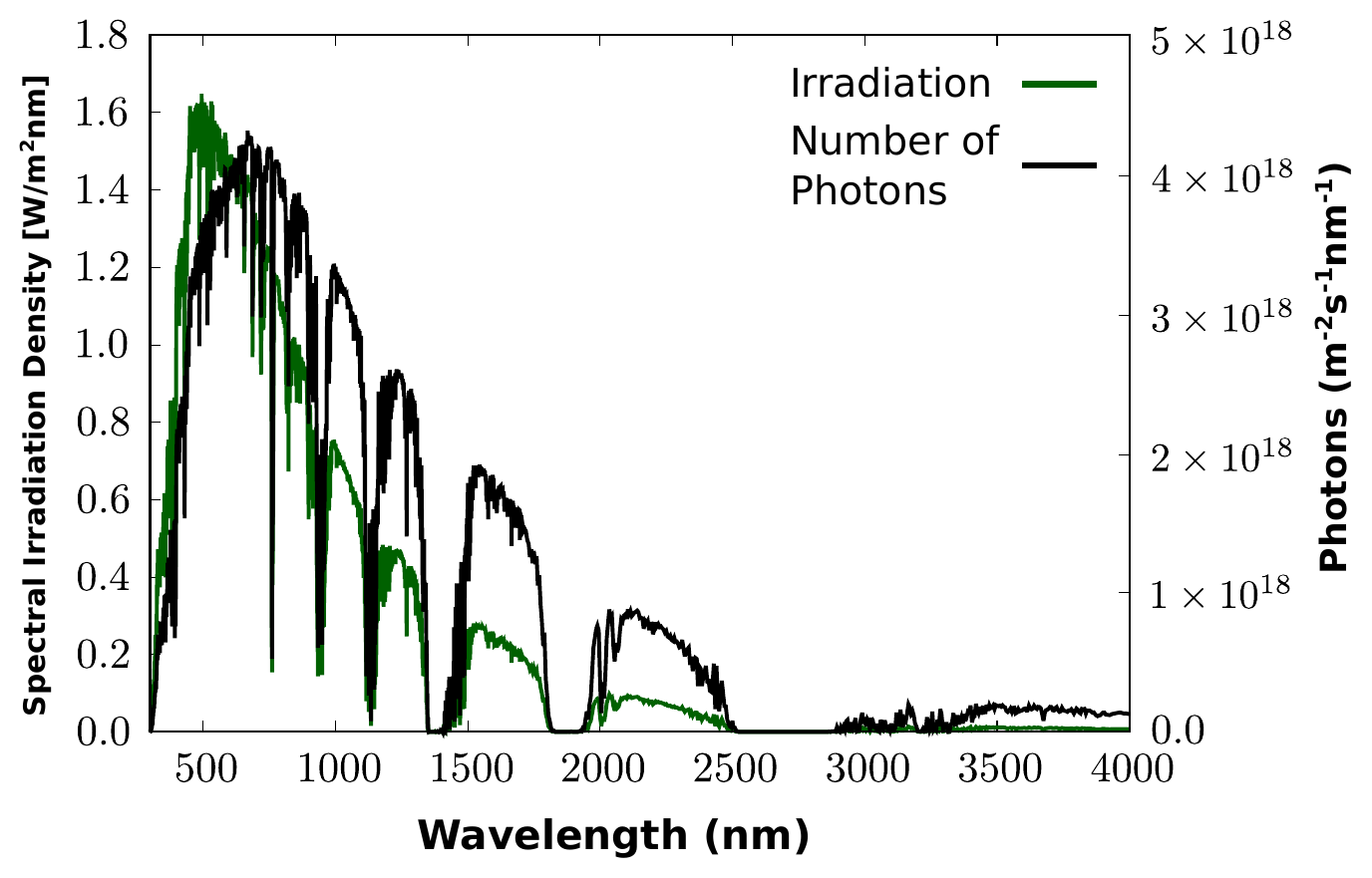}
\caption{Sun irradiance (green) and number of photons (black) as a function of wavelength.}
\label{Fig4}
\end{figure}

\begin{figure}[htp]
\centering
\includegraphics[scale=0.5]{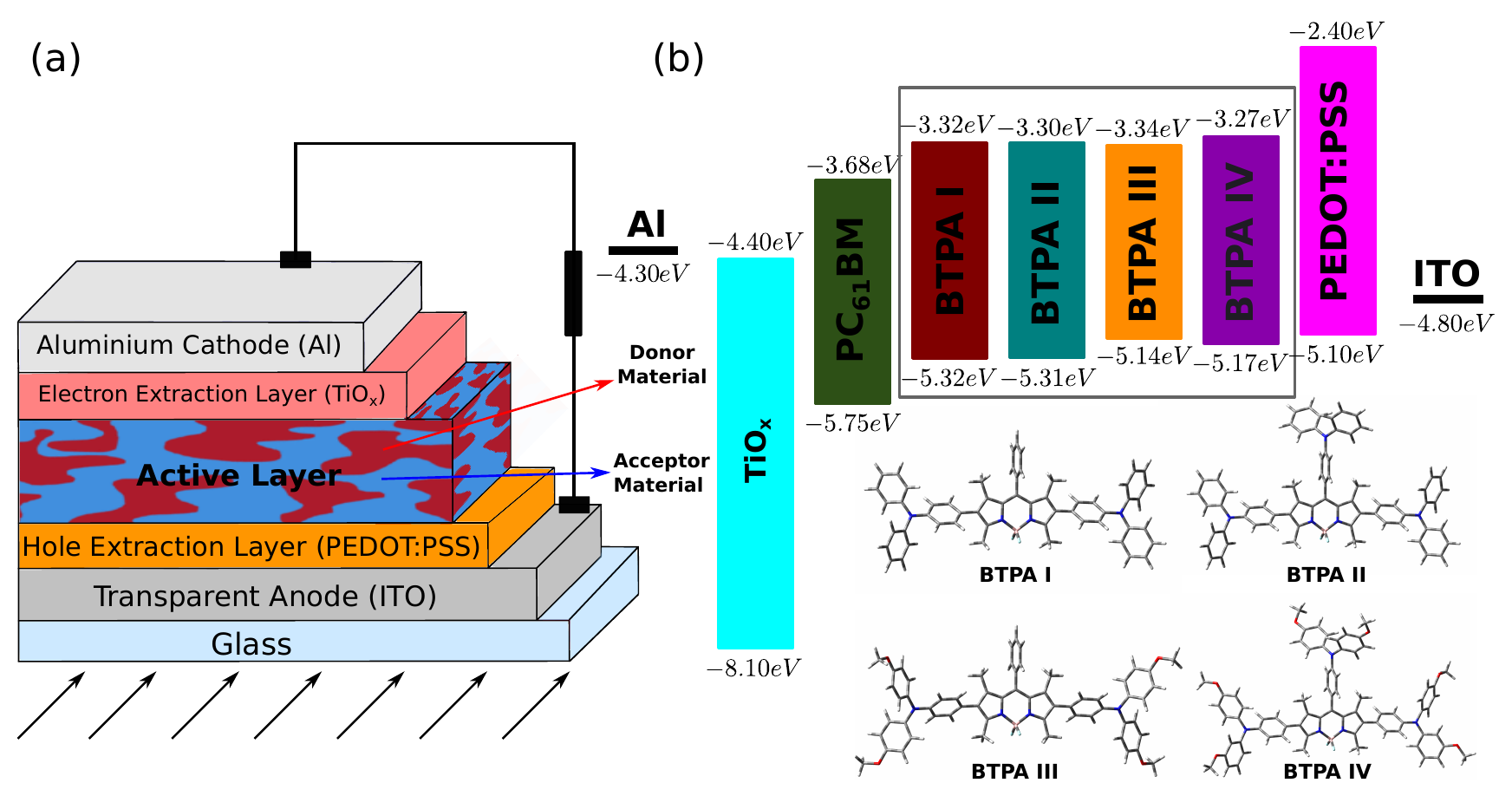}
\caption{(a) Schematic diagram of BHJ OSC and, (b) energy band diagram to represent the energy levels.}
\label{Fig5}
\end{figure}

The sun emits an enormous amount of energy every second in the form of irradiation, reaching Earth with an energy density of about $1366$ Wm$^{-2}$ just outside the atmosphere, although some energy will be lost after reflection and absorption by the atmosphere~\sugb{\cite{armaroli2007}}. Assuming that in a BHJ organic solar cell (Fig.~\sugb{\ref{Fig5}}) the electron-acceptor system in the active layer defines the optical space of the compound, the density of absorbed photons can be calculated as well as the density of power combining the absorption spectrum of the molecular system with the spectrum of the sun. The typical spectrum of light incident on the Earth's surface is given by the ASTM G159~\sugb{\cite{American}} standard and is called Air Mass 1.5 (AM1.5), as shown in Fig.~\sugb{\ref{Fig4}}, which indicates the energy density and photon flux with respect to wavelength. 
The AM1.5G, is the general reference for solar cell characterization~\sugb{\cite {shrotriya2016}}, accumulates an integrated power density of 1000~Wm$^{-2}$ (100~mWcm$^{-2}$) and an integrated photon flux of $4.31\times 10^{21}$~s$^{-1}$m$^{-2}$, distributed over a wide range of wavelengths ($280$--$4000$~nm). Therefore, to efficiently collect solar energy, the absorption spectra of CSOs must have a large overlap with the solar spectrum in this region. Thus, the first criteria in the design of new materials for the active layer of an OSC is its high efficiency in absorbing solar energy throughout the solar spectrum.
As mentioned above, the band gap determines the absorption limit for a particular molecular system. As such, much effort has gone into the research community to improve the absorption characteristics of organic solar cells by fine-tuning the absorption characteristics. With a low bandgap and wide absorption band, a molecular system can absorb more photons, which will increase the $J_{sc}$, all other things being equal. However, further reducing the bandgap of a molecular system will lead to a decrease in $V_{oc}$, which is believed to correlate with the difference between the highest occupied molecular orbital (HOMO) of the donor materials and the lowest unoccupied molecular orbital level (LUMO) of the acceptor materials.
\begin{table}[htp] 
\begin{center}
\centering
\caption{Parameter values of the studied molecular compounds: Energy of the exciton driving force $\Delta E$, open circuit voltage ($V_{oc}$), short-circuit current density ($J_{sc}$), fill factor ($FF$), and power conversion efficiency ($PCE$).}
\scalebox{0.81}{
\begin{threeparttable}
\begin{tabular}{ccccccc} 
\hline
\hline
&&&&&&\\
 \centering \textbf{\textit{System}} & $\Delta$\textbf{\textit{E}} (eV) & \textbf{\textit{V}}$_{oc}$ (V) & \textbf{\textit{J}}$_{sc}$ (mAcm$^{-2}$) & \textbf{\textit{FF}} & \textbf{\textit{PCE}} (\%) \\
&&&&&&\\ 
\hline
\hline
&&&&&&\\
\textbf{\textbf{BTPA~I}}   & 0.360 & 1.342 & 11.614 & 0.841 & 13.113 \\ 
\textbf{\textbf{BTPA~II}}  & 0.370 & 1.336 & 11.489 & 0.840 & 12.897 \\  
\textbf{\textbf{BTPA~III}} & 0.340 & 1.166 & 15.608 & 0.824 & 14.988 \\
\textbf{\textbf{BTPA~IV}}  & 0.408 & 1.193 & 13.553 & 0.827 & 13.366 \\
&&&&&&\\
\hline
\hline
\end{tabular}
\end{threeparttable}}
\label{Tab3}
\end{center}
\end{table}
Therefore, in this study the Scharbers model is implemented to study the photovoltaic properties, as $V_{oc}$ and $J_{sc}$ of hole transportin materials \textbf{\textit{BTPA~I--IV}}, with the electron transporting material (\textbf{\textit{PC$_{61}$BM}}) in the active layer of a BHJ organic solar cell with a structure as shown in Fig.~\sugb{\ref{Fig5}} and estimate the energy conversion capacity of the systems. Although the $PCEs$ of organic photovoltaic (OPV) cells have been improved to some extent, it is still far from commercial application compared to the 25.2~\% $PCE$ of lead halide perovkite-based thin film solar cells~\sugb{\cite{xiao2017}}. The performance of the OPV needs to be further improved. Basically, the $PCE$ of a photovoltaic cell can be expressed as:

\begin{equation}
PCE=\frac{V_{oc}J_{sc}FF}{P_{in}}
\label{Ecu4}
\end{equation}

where $J_{sc}$ means the short-circuit current density, $V_{oc}$ denotes the open-circuit photovoltage, $FF$ is the fill factor, and $P_{in}$ represents the power of incident light~\sugb{\citep{berube2013}}. Clearly, to get a better PCE, none of these four components can be ignored in the material design. Generally, $V_{oc}$ is related to the energy difference between the HOMO of the donor material and the LUMO of the acceptor material. It is also affected by the rates of carrier generation and recombination, the presence of energetic tail states or trap states~\sugb{\cite{lange2013, sweetnam2014}}. The values for $V_{oc}$ are shown in Table~\sugb{\ref{Tab3}}, calculated using the empirical equation summarized by Scharber et al.~\sugb{\cite{scharber2006, scharber2013}}, which can be expressed as

\begin{equation}
V_{oc}=\frac{1}{e}\big(\vert E_{H}^{Donor}\vert-\vert E_{L}^{PC_{61}BM}\vert\big)-0.3 V
\label{Ecu5}
\end{equation}

where $e$ is the elementary charge, the value of $0.3$~V is a typical loss found in bulk heterojunction solar cells, and using $-3.69$~eV for $E_L$ of \textbf{\textit{PC$_{61}$BM}} lowest occupied molecular orbital (LUMO) energy.
\medskip

In operational solar cells additional recombination paths decrease the value of $V_{oc}$ affecting the $PCE$ of OPVs because they must operate at a voltage lower than the $V_{oc}$. The $V_{oc}$ decreasing implies a power reduction ($J_{sc}\cdot V_{oc}$) given by the fill factor ($FF$). Based on these considerations, the ideal parameters for a $\pi$--conjugated-system-\textbf{\textit{PC$_{61}$BM}} device are shown in Table~\sugb{\ref{Tab3}}. The $V_{oc}$ values of the \textbf{\textit{BTPA~I--IV}} systems are in the range of $1.166$--$1.342$~V, which indicates values greater than $1.0$~V efficiently increasing the value of $V_{oc}$, and furthermore, since $\Delta E$ shows values greater than $0.3$~eV, this suggests that the transfer of photoexcited eletrons is easy from the material carrier hole to the acceptor\textbf{\textit{PC$_{61}$BM}}. Therefore, devices constructed from \textbf{\textit{BTPA~I--IV}} and \textbf{\textit{PC$_{61}$BM}} material can exhibit high-efficiency electron injection, which implies that molecular systems \textbf{\textit{BTPA~I--IV}} could become good potential candidates for organic photovoltaic devices.
\medskip

Given that the $PCEs$ are estimated from the values for the open circuit voltage $V_{oc}$, the short-circuit current $J_{sc}$, and the fill factor $FF$ of the OSCs~\sugb{\cite{guo2012}}, to achieve high efficiencies, an ideal donor material should have a low energy gap and a deep HOMO energy level (thus increasing $V_{oc}$). Additionally, a high hole mobility is also crucial for the carrier transport to improve $J_{sc}$ and $FF$,Therefore we can calculate the potential $FF$ of the cells that could be achieved if there were no additional load transport losses. Which can be described using Green's approximation \sugb{\cite{green1982,   stolterfoht2017}}

\begin{equation}
FF=\frac{\nu_{oc}-ln(\nu_{_{oc}}+0.72)}{\nu_{oc}+1},
\label{Ecu6}
\end{equation}

\noindent with $\nu_{oc}=\frac{V_{oc}}{n\kappa_{_B}T}$. The parameter values corresponding to the \textbf{\textit{BTPA~I}}, \textbf{\textit{BTPA~II}}, \textbf{\textit{BTPA~III}}, and \textbf{\textit{BTPA~IV}} compounds are shown in Table~\sugb{\ref{Tab3}}. The fill factor oscillates between $0.827$ to $0.841$ in THF, in agreement with the reported BHJ solar cells values for different polymers as electron-donor, and \textbf{\textit{PC$_{61}$BM}} as electron-acceptor material~\sugb{\cite{huo2015,trukhanov2015}}.
\medskip

On the other hand, it is known that the short-circuit current density ($J_{sc}$) is another key parameter to evaluate the performance of BHJ solar cells because it is directly associated with the energy conversion efficiency. For maximum efficiency, the spectral shape of sunlight must be taken into account, since the intensity and spectral shape depend on various circumstances, eg angle above the horizon, a standard power spectrum is defined $P_{AM1.5}(\lambda)$~\sugb{\cite{armaoli2007, shrotriya2016, American}} for an angle above the horizon of $45$ degrees. So $J_{sc}$ can be determined by the following expression~\sugb{\cite{dennler2009, el2015}},

\begin{equation}
J_{sc}=\frac{q}{hc}\int_{0}^{E_g^D}EQE(\lambda)P_{AM1.5G}(\lambda) \lambda d\lambda
\label{Ecu7}
\end{equation}

where $h$ is Plank’s constant, $c$ is the speed of light in vacuum, and $E_g^D$ is the bad gap of the hole transporting material in the active layer of the device and $EQE(\lambda)$ is the external quantum efficiency, which for this work is assumed with a value of $80\%$~\sugb{\cite{scharber2013,jagadamma2015,li2015}}. The values of $J_{sc}$ and $PCE$ for the systems \textbf{\textit{BTPA~I--IV}} are shown in Table~\sugb{\ref{Tab3}}, in which it is observed that the values of $J_{sc}$ are in a range of $11.489$--$15.608$~mAcm$^{-2}$, finding the highest value for the molecular system \textit{\textbf{BTPA~III}}, which also has the highest value of $PCE$ around $14.988~\%$. This study shows with respect to the $PCE$ of the systems considered and taking as a reference the \textbf{\textit{BTPA~I}} system, that the inclusion of the methoxyl group (--O-CH$_3$) in different positions to generate the system \textbf{\textit{BTPA~III}}, favors the increase of the $PCE$ by $9.57~\%$. However, simply including triarylamine (\textbf{\textit{BTPA~II}}) disadvantages the $PCE$, presenting a reduction of $1.06~\%$ compared to the \textbf{\textit{BTPA~I}} system, While the inclusion of triarylamine and the methoxy group in 6 different positions increases the value of $PCE$, we only have an increase of $1.69~\%$ compared to the value presented by the molecular system \textbf{\textit{BTPA~I}}. Which shows that the \textbf{\textit{BTPA~III}} molecular system presents the greatest potential for use as photovoltaic materials in BHJ solar cell.

\section{Conclusions}\label{Conclusions}

We have studied four D-$\pi$-A molecular systems that contain DODIPY, triarylamine, methoxy fragment in their structure. The geometric, electronic, optical and photovoltaic properties of the designed systems were investigated using DFT calculations with the theory levels B3LYP/6-31G(d,p) and TD-DFT/CAM-B3LYP/6-31G(d,p). The estimation of the energy of the HOMO-LUMO boundary molecular orbitals showed an important correlation with the available experimental data~\sugb{\cite{ortiz2019}}. The energy of the exciton driving force ($\Delta E$) of all molecular systems (\textbf{\textit{BTPA~I--IV}}) has a value greater than $0.3$~eV, which can guarantee efficient division of excitons. The simulated UV/VIS absorption spectra show a similar profile for all systems, which present a strong main band between $536.910$~nm and $543.110$~nm. In general, molecular systems \textbf{\textit{BTPA~III}} and \textbf{\textit{BTPA~IV}} have the best photovoltaic properties compared to derivatives \textbf{\textit{BTPA~I}} and \textbf{\textit{BTPA~II}}, where low values were observed in the gap energy ($E_g$) ($1.806$--$1.901$~eV for \textbf{\textit{BTPA~III}} and \textbf{\textit{BTPA~IV}}, respectively) and high values in the short-circuit photocurrent density ($J_{sc}$) ($15.608$ and $13.553$~mAcm$^{-2}$) and in the photoelectric conversion efficiency ($PCE$) ($14.998$--$13.366~\%$), therefore, these derivatives can potentially be used in organic solar cells (OSCs) of volume heterojunction (BHJ) and also the great importance of the theoretical study of orbitals.

\section*{Acknowledgements}

D. M. U. acknowledge J. Vélez for helpful with computational resources. This work was supported by the Colombian Science, Technology and Innovation Fund-General Royalties System (Fondo CTeI-Sistema General de Regal\'ias) under contract BPIN 2013000100007.

\bibliographystyle{ieeetr}
\bibliography{Bibliography}

\begin{thebibliography}{10}

\bibitem{you2013}
J.~You, L.~Dou, K.~Yoshimura, T.~Kato, K.~Ohya, T.~Moriarty, K.~Emery, C.-C.
  Chen, J.~Gao, G.~Li, {\em et~al.}, ``A polymer tandem solar cell with 10.6\%
  power conversion efficiency,'' {\em Nature communications}, vol.~4, no.~1,
  pp.~1--10, 2013.

\bibitem{huang2014}
Y.~Huang, E.~J. Kramer, A.~J. Heeger, and G.~C. Bazan, ``Bulk heterojunction
  solar cells: morphology and performance relationships,'' {\em Chemical
  reviews}, vol.~114, no.~14, pp.~7006--7043, 2014.

\bibitem{aruna2015}
P.~Aruna, K.~Suresh, and C.~Joseph, ``Effect of fullerene doping on the
  electrical properties of p3ht/pcbm layers,'' {\em Materials Science in
  Semiconductor Processing}, vol.~36, pp.~7--12, 2015.

\bibitem{wang2016}
M.~Wang, D.~Cai, Z.~Yin, S.-C. Chen, C.-F. Du, and Q.~Zheng,
  ``Asymmetric-indenothiophene-based copolymers for bulk heterojunction solar
  cells with 9.14\% efficiency,'' {\em Advanced Materials}, vol.~28, no.~17,
  pp.~3359--3365, 2016.

\bibitem{xia2017}
Y.~Xia, J.~Fang, P.~Li, B.~Zhang, H.~Yao, J.~Chen, J.~Ding, and J.~Ouyang,
  ``Solution-processed highly superparamagnetic and conductive pedot: Pss/fe3o4
  nanocomposite films with high transparency and high mechanical flexibility,''
  {\em ACS Applied Materials \& Interfaces}, vol.~9, no.~22, pp.~19001--19010,
  2017.

\bibitem{yang2015}
D.~Yang, L.~Yang, Y.~Huang, Y.~Jiao, T.~Igarashi, Y.~Chen, Z.~Lu, X.~Pu,
  H.~Sasabe, and J.~Kido, ``Asymmetrical squaraines bearing
  fluorine-substituted indoline moieties for high-performance
  solution-processed small-molecule organic solar cells,'' {\em ACS Applied
  Materials \& Interfaces}, vol.~7, no.~24, pp.~13675--13684, 2015.

\bibitem{ni2015}
W.~Ni, X.~Wan, M.~Li, Y.~Wang, and Y.~Chen, ``A--d--a small molecules for
  solution-processed organic photovoltaic cells,'' {\em Chemical
  communications}, vol.~51, no.~24, pp.~4936--4950, 2015.

\bibitem{collins2017}
S.~D. Collins, N.~A. Ran, M.~C. Heiber, and T.-Q. Nguyen, ``Small is powerful:
  recent progress in solution-processed small molecule solar cells,'' {\em
  Advanced Energy Materials}, vol.~7, no.~10, p.~1602242, 2017.

\bibitem{zhang2018}
Y.~Zhang, B.~Kan, Y.~Sun, Y.~Wang, R.~Xia, X.~Ke, Y.-Q.-Q. Yi, C.~Li, H.-L.
  Yip, X.~Wan, {\em et~al.}, ``Nonfullerene tandem organic solar cells with
  high performance of 14.11\%,'' {\em Advanced Materials}, vol.~30, no.~18,
  p.~1707508, 2018.

\bibitem{yan2018}
C.~Yan, S.~Barlow, Z.~Wang, H.~Yan, A.~K.-Y. Jen, S.~R. Marder, and X.~Zhan,
  ``Non-fullerene acceptors for organic solar cells,'' {\em Nature Reviews
  Materials}, vol.~3, no.~3, pp.~1--19, 2018.

\bibitem{zhang2012}
L.~Zhang, K.~Pei, M.~Yu, Y.~Huang, H.~Zhao, M.~Zeng, Y.~Wang, and J.~Gao,
  ``Theoretical investigations on donor--acceptor conjugated copolymers based
  on naphtho [1, 2-c: 5, 6-c] bis [1, 2, 5] thiadiazole for organic solar cell
  applications,'' {\em The Journal of Physical Chemistry C}, vol.~116, no.~50,
  pp.~26154--26161, 2012.

\bibitem{kim2013}
B.-G. Kim, X.~Ma, C.~Chen, Y.~Ie, E.~W. Coir, H.~Hashemi, Y.~Aso, P.~F. Green,
  J.~Kieffer, and J.~Kim, ``Energy level modulation of homo, lumo, and band-gap
  in conjugated polymers for organic photovoltaic applications,'' {\em Advanced
  Functional Materials}, vol.~23, no.~4, pp.~439--445, 2013.

\bibitem{Belghiti2012}
H.~M. B. S. B.~M. Belghiti~N., Bennani~M., ``Xxxxx,'' {\em Afr. J. Pure. Appl.
  Chem.}, vol.~6, no.~164, p.~XX, 2012.

\bibitem{berube2013}
N.~B{\'e}rub{\'e}, V.~Gosselin, J.~Gaudreau, and M.~Cote, ``Designing polymers
  for photovoltaic applications using ab initio calculations,'' {\em The
  Journal of Physical Chemistry C}, vol.~117, no.~16, pp.~7964--7972, 2013.

\bibitem{scharber2006}
M.~C. Scharber, D.~M{\"u}hlbacher, M.~Koppe, P.~Denk, C.~Waldauf, A.~J. Heeger,
  and C.~J. Brabec, ``Design rules for donors in bulk-heterojunction solar
  cells—towards 10\% energy-conversion efficiency,'' {\em Advanced
  materials}, vol.~18, no.~6, pp.~789--794, 2006.

\bibitem{dennler2009}
G.~Dennler, M.~C. Scharber, and C.~J. Brabec, ``Polymer-fullerene
  bulk-heterojunction solar cells,'' {\em Advanced materials}, vol.~21, no.~13,
  pp.~1323--1338, 2009.

\bibitem{azazi2011}
A.~Azazi, A.~Mabrouk, and K.~Alimi, ``Theoretical investigation on the
  photophysical properties of low-band-gap copolymers for photovoltaic
  devices,'' {\em Computational and Theoretical Chemistry}, vol.~978, no.~1-3,
  pp.~7--15, 2011.

\bibitem{scharber2013}
M.~C. Scharber and N.~S. Sariciftci, ``Efficiency of bulk-heterojunction
  organic solar cells,'' {\em Progress in polymer science}, vol.~38, no.~12,
  pp.~1929--1940, 2013.

\bibitem{wang2014}
D.~Wang, X.~Zhang, W.~Ding, X.~Zhao, and Z.~Geng, ``Density functional theory
  design and characterization of d--a--a type electron donors with narrow band
  gap for small-molecule organic solar cells,'' {\em Computational and
  Theoretical Chemistry}, vol.~1029, pp.~68--78, 2014.

\bibitem{scharber2016}
M.~C. Scharber, ``On the efficiency limit of conjugated polymer:
  fullerene-based bulk heterojunction solar cells,'' {\em Advanced Materials},
  vol.~28, no.~10, pp.~1994--2001, 2016.

\bibitem{taouali2019}
W.~Taouali, M.~E. Casida, S.~Znaidia, and K.~Alimi, ``Rational design of (da)
  copolymers towards high efficiency organic solar cells: Dft and td-dft
  study,'' {\em Journal of Molecular Graphics and Modelling}, vol.~89,
  pp.~139--146, 2019.

\bibitem{siddiqui2019}
A.~Siddiqui, M.~Keshtov, G.~D. Sharma, S.~P. Singh, {\em et~al.}, ``New indolo
  carbazole-based non-fullerene n-type semiconductors for organic solar cell
  applications,'' {\em Journal of Materials Chemistry C}, vol.~7, no.~3,
  pp.~543--552, 2019.

\bibitem{padula2019}
D.~Padula, J.~D. Simpson, and A.~Troisi, ``Combining electronic and structural
  features in machine learning models to predict organic solar cells
  properties,'' {\em Materials Horizons}, vol.~6, no.~2, pp.~343--349, 2019.

\bibitem{raftani2020}
M.~Raftani, T.~Abram, M.~Bennani, and M.~Bouachrine, ``Theoretical study of new
  conjugated compounds with a low bandgap for bulk heterojunction solar cells:
  Dft and td-dft study,'' {\em Results in Chemistry}, p.~100040, 2020.

\bibitem{afzal2020}
Z.~Afzal, R.~Hussain, M.~U. Khan, M.~Khalid, J.~Iqbal, M.~U. Alvi, M.~Adnan,
  M.~Ahmed, M.~Y. Mehboob, M.~Hussain, {\em et~al.}, ``Designing
  indenothiophene-based acceptor materials with efficient photovoltaic
  parameters for fullerene-free organic solar cells.,'' {\em Journal of
  Molecular Modeling}, vol.~26, no.~6, pp.~137--137, 2020.

\bibitem{hachi2020}
M.~Hachi, A.~Slimi, A.~Fitri, S.~ElKhattabi, A.~T. Benjelloun, M.~Benzakour,
  and M.~Mcharfi, ``New small organic molecules based on thieno [2, 3-b] indole
  for efficient bulk heterojunction organic solar cells: a computational
  study,'' {\em Molecular Physics}, vol.~118, no.~8, p.~e1662956, 2020.

\bibitem{mahdavifar2020}
Z.~Mahdavifar, S.~Tajdinan, and E.~Shakerzadeh, ``Quantum mechanical
  calculations of photovoltaic and photoelectronic properties of
  oligoselenophene/fullerene bhj solar cells,'' {\em Inorganic Chemistry
  Research}, pp.~94--102, 2020.

\bibitem{ortiz2019}
A.~Ortiz, ``Triarylamine-bodipy derivatives: A promising building block as hole
  transporting materials for efficient perovskite solar cells,'' {\em Dyes and
  Pigments}, vol.~171, p.~107690, 2019.

\bibitem{frisch2009}
M.~Frisch, G.~Trucks, H.~B. Schlegel, G.~E. Scuseria, M.~A. Robb, J.~R.
  Cheeseman, G.~Scalmani, V.~Barone, B.~Mennucci, G.~Petersson, and at~al.,
  ``Gaussian 09, revision d. 01,'' 2009.

\bibitem{becke1993}
A.~D. Becke, ``Density-functional thermochemistry {III}. the role of exact
  exchange,'' {\em J. Chem. Phys.}, vol.~98, no.~7, pp.~5648--5652, 1993.

\bibitem{madrid2018}
D.~Madrid-{\'U}suga, C.~A. Melo-Luna, A.~Insuasty, A.~Ortiz, and J.~H. Reina,
  ``Optical and electronic properties of molecular systems derived from
  rhodanine,'' {\em The Journal of Physical Chemistry A}, vol.~122, no.~43,
  pp.~8469--8476, 2018.

\bibitem{calderon2020}
K.~Calderon-Cerquera, A.~Parra, D.~Madrid, A.~Cabrera-Espinoza, C.~A.
  Melo-Luna, J.~H. Reina, B.~Insuasty, and A.~Ortiz, ``Synthesis,
  characterization and photophysics of novel bodipy linked to dumbbell systems
  based on fullerene [60] pyrrolidine and fullerene [60] isoxazoline,'' {\em
  Dyes and Pigments}, p.~108752, 2020.

\bibitem{ganji2015}
M.~D. Ganji, S.~Hosseini-Khah, and Z.~Amini-Tabar, ``Theoretical insight into
  hydrogen adsorption onto graphene: a first-principles {B3LYP-D3} study,''
  {\em Phys. Chem. Chem. Phys.}, vol.~17, no.~4, pp.~2504--2511, 2015.

\bibitem{ganji2016}
M.~D. Ganji, M.~Tajbakhsh, M.~Kariminasab, and H.~Alinezhad, ``Tuning the
  {LUMO} level of organic photovoltaic solar cells by conjugately fusing
  graphene flake: A {DFT-B3LYP} study,'' {\em Physica E Low Dimens. Syst.
  Nanostruct}, vol.~81, pp.~108--115, 2016.

\bibitem{takano2005}
Y.~Takano and K.~Houk, ``Benchmarking the conductor-like polarizable continuum
  model ({CPCM}) for aqueous solvation free energies of neutral and ionic
  organic molecules,'' {\em J. Chem. Theory Comput.}, vol.~1, no.~1,
  pp.~70--77, 2005.

\bibitem{chiu2016}
K.~Y. Chiu, V.~Govindan, L.-C. Lin, S.-H. Huang, J.-C. Hu, K.-M. Lee, H.-H.~G.
  Tsai, S.-H. Chang, and C.-G. Wu, ``{DPP} containing {D--$\pi$--A} organic
  dyes toward highly efficient dye-sensitized solar cells,'' {\em Dyes and
  Pigm.}, vol.~125, pp.~27--35, 2016.

\bibitem{zhang2013}
Z.-L. Zhang, L.-Y. Zou, A.-M. Ren, Y.-F. Liu, J.-K. Feng, and C.-C. Sun,
  ``Theoretical studies on the electronic structures and optical properties of
  star-shaped triazatruxene/heterofluorene co-polymers,'' {\em Dyes and
  Pigments}, vol.~96, no.~2, pp.~349--363, 2013.

\bibitem{jagadamma2015}
L.~K. Jagadamma, M.~Al-Senani, A.~El-Labban, I.~Gereige, G.~O. Ngongang~Ndjawa,
  J.~C. Faria, T.~Kim, K.~Zhao, F.~Cruciani, D.~H. Anjum, {\em et~al.},
  ``Polymer solar cells with efficiency> 10\% enabled via a facile
  solution-processed al-doped zno electron transporting layer,'' {\em Advanced
  Energy Materials}, vol.~5, no.~12, p.~1500204, 2015.

\bibitem{brabec2011}
C.~J. Brabec, M.~Heeney, I.~McCulloch, and J.~Nelson, ``Influence of blend
  microstructure on bulk heterojunction organic photovoltaic performance,''
  {\em Chemical Society Reviews}, vol.~40, no.~3, pp.~1185--1199, 2011.

\bibitem{jo2012}
M.~Y. Jo, S.~J. Park, T.~Park, Y.~S. Won, and J.~H. Kim, ``Relationship between
  homo energy level and open circuit voltage of polymer solar cells,'' {\em
  Organic Electronics}, vol.~13, no.~10, pp.~2185--2191, 2012.

\bibitem{wang2014r}
D.~Wang, W.~Ding, Z.~Geng, L.~Wang, Y.~Geng, Z.~Su, and H.~Yu, ``Rational
  design and characterization of high-efficiency planar a--$\pi$--d--$\pi$--a
  type electron donors in small molecule organic solar cells: A quantum
  chemical approach,'' {\em Materials Chemistry and Physics}, vol.~145, no.~3,
  pp.~387--396, 2014.

\bibitem{walker2013}
B.~Walker, J.~Liu, C.~Kim, G.~C. Welch, J.~K. Park, J.~Lin, P.~Zalar, C.~M.
  Proctor, J.~H. Seo, G.~C. Bazan, {\em et~al.}, ``Optimization of energy
  levels by molecular design: evaluation of bis-diketopyrrolopyrrole molecular
  donor materials for bulk heterojunction solar cells,'' {\em Energy \&
  Environmental Science}, vol.~6, no.~3, pp.~952--962, 2013.

\bibitem{duan2013}
Y.-A. Duan, Y.~Geng, H.-B. Li, J.-L. Jin, Y.~Wu, and Z.-M. Su, ``Theoretical
  characterization and design of small molecule donor material containing
  naphthodithiophene central unit for efficient organic solar cells,'' {\em
  Journal of computational chemistry}, vol.~34, no.~19, pp.~1611--1619, 2013.

\bibitem{dkhissi2011}
A.~Dkhissi, ``Excitons in organic semiconductors,'' {\em Synthetic metals},
  vol.~161, no.~13-14, pp.~1441--1443, 2011.

\bibitem{kose2012}
M.~E. Kose, ``Evaluation of acceptor strength in thiophene coupled
  donor--acceptor chromophores for optimal design of organic photovoltaic
  materials,'' {\em The Journal of Physical Chemistry A}, vol.~116, no.~51,
  pp.~12503--12509, 2012.

\bibitem{clarke2010}
T.~M. Clarke and J.~R. Durrant, ``Charge photogeneration in organic solar
  cells,'' {\em Chemical reviews}, vol.~110, no.~11, pp.~6736--6767, 2010.

\bibitem{renuga2014}
S.~Renuga and S.~Muthu, ``Molecular structure, normal coordinate analysis,
  harmonic vibrational frequencies, nbo, homo--lumo analysis and detonation
  properties of (s)-2-(2-oxopyrrolidin-1-yl) butanamide by density functional
  methods,'' {\em Spectrochimica Acta Part A: Molecular and Biomolecular
  Spectroscopy}, vol.~118, pp.~702--715, 2014.

\bibitem{dhas2010}
D.~A. Dhas, I.~H. Joe, S.~Roy, and T.~Freeda, ``Dft computations and
  spectroscopic analysis of a pesticide: Chlorothalonil,'' {\em Spectrochimica
  Acta Part A: Molecular and Biomolecular Spectroscopy}, vol.~77, no.~1,
  pp.~36--44, 2010.

\bibitem{sangeetha2016}
M.~Sangeetha and R.~Mathammal, ``A complete synergy on the experimental and
  theoretical study of the pyridine derivatives--2-hydroxy-5-nitropyridine and
  2-chloro-5-nitropyridine,'' {\em Journal of Molecular Structure}, vol.~1117,
  pp.~121--134, 2016.

\bibitem{armaroli2007}
N.~Armaroli and V.~Balzani, ``The future of energy supply: challenges and
  opportunities,'' {\em Angewandte Chemie International Edition}, vol.~46,
  no.~1-2, pp.~52--66, 2007.

\bibitem{American}
{\em \textit{American Society for Testing and Materials (ASTM) Standard G159,
  West Conshoken, PA, USA. Source:
  http://rredc.nrel.gov/solar/spectra/am1.5/.}}
\newblock http://rredc.nrel.gov/solar/spectra/am1.5/.

\bibitem{shrotriya2016}
V.~Shrotriya, G.~Li, Y.~Yao, T.~Moriarty, K.~Emery, and Y.~Yang, ``Accurate
  measurement and characterization of organic solar cells,'' {\em Advanced
  functional materials}, vol.~16, no.~15, pp.~2016--2023, 2016.

\bibitem{xiao2017}
Z.~Xiao and Y.~Yan, ``Progress in theoretical study of metal halide perovskite
  solar cell materials,'' {\em Advanced Energy Materials}, vol.~7, no.~22,
  p.~1701136, 2017.

\bibitem{lange2013}
I.~Lange, J.~Kniepert, P.~Pingel, I.~Dumsch, S.~Allard, S.~Janietz, U.~Scherf,
  and D.~Neher, ``Correlation between the open circuit voltage and the
  energetics of organic bulk heterojunction solar cells,'' {\em J. Phys. Chem.
  Lett.}, vol.~4, no.~22, pp.~3865--3871, 2013.

\bibitem{sweetnam2014}
S.~Sweetnam, K.~R. Graham, G.~O. Ngongang~Ndjawa, T.~Heumüller, J.~A. Bartelt,
  T.~M. Burke, W.~Li, W.~You, A.~Amassian, and M.~D. McGehee,
  ``Characterization of the polymer energy landscape in polymer: fullerene bulk
  heterojunctions with pure and mixed phases,'' {\em Journal of the American
  Chemical Society}, vol.~136, no.~40, pp.~14078--14088, 2014.

\bibitem{guo2012}
X.~Guo, M.~Zhang, L.~Huo, C.~Cui, Y.~Wu, J.~Hou, and Y.~Li, ``Poly(thieno [3,
  2-b]thiophene-alt-bithiazole): {A--D--A} copolymer donor showing improved
  photovoltaic performance with indene-c$_{60}$ bisadduct acceptor,'' {\em
  Macromolecules}, vol.~45, no.~17, pp.~6930--6937, 2012.

\bibitem{green1982}
M.~A. Green, ``Accuracy of analytical expressions for solar cell fill
  factors,'' {\em Sol. Cells}, vol.~7, no.~3, pp.~337--340, 1982.

\bibitem{stolterfoht2017}
M.~Stolterfoht, C.~M. Wolff, Y.~Amir, A.~Paulke, L.~Perdig{\'o}n-Toro,
  P.~Caprioglio, and D.~Neher, ``Approaching the fill factor shockley--queisser
  limit in stable, dopant-free triple cation perovskite solar cells,'' {\em
  Energy \& Environmental Science}, vol.~10, no.~6, pp.~1530--1539, 2017.

\bibitem{huo2015}
L.~Huo, T.~Liu, X.~Sun, Y.~Cai, A.~J. Heeger, and Y.~Sun, ``Single-junction
  organic solar cells based on a novel wide-bandgap polymer with efficiency of
  9.7\%,'' {\em Adv. Mater.}, vol.~27, no.~18, pp.~2938--2944, 2015.

\bibitem{trukhanov2015}
V.~A. Trukhanov, V.~V. Bruevich, and D.~Y. Paraschuk, ``Fill factor in organic
  solar cells can exceed the shockley-queisser limit,'' {\em Sci. Rep.},
  vol.~5, p.~11478, 2015.

\bibitem{el2015}
A.~El~Assyry, R.~Jdaa, B.~Benali, M.~Addou, and A.~Zarrouk, ``Optical and
  photovoltaic properties of new quinoxalin-2 (1h)-one-based da organic dyes
  for efficient dye-sensitized solar cell using dft,'' {\em J. Mater. Environ.
  Sci}, vol.~6, pp.~2612--2623, 2015.

\bibitem{li2015}
N.~Li and C.~J. Brabec, ``Air-processed polymer tandem solar cells with power
  conversion efficiency exceeding 10\%,'' {\em Energy \& Environmental
  Science}, vol.~8, no.~10, pp.~2902--2909, 2015.

\end{thebibliography}
\end{document}